\documentclass[12pt]{article}
\pdfoutput=1  
\usepackage{hyperref}
\usepackage[pdftex]{graphicx}
\usepackage{amsmath,amssymb}
\usepackage{cancel}
\usepackage{appendix}
\usepackage{mathtools}
\usepackage[mathscr]{eucal}
\newcommand{\cO}{{\mathcal{O}}}

\begin{document}

\begin{titlepage}


\begin{center}
\Large \bf \textbf{Asymptotically Nonrelativistic \\
String Backgrounds}
\end{center}

\begin{center}
Daniel \'Avila$^{\ast}$,
Alberto G\"uijosa$^{\ast}$
and Rafael Olmedo$^{\ast}$

\vspace{0.2cm}
$^{\ast}\,$Departamento de F\'{\i}sica de Altas Energ\'{\i}as, Instituto de Ciencias Nucleares, \\
Universidad Nacional Aut\'onoma de M\'exico,
\\ Apartado Postal 70-543, CdMx 04510, M\'exico\\
 \vspace{0.2cm}
\vspace{0.2cm}
{\tt daniel.avila@correo.nucleares.unam.mx, alberto@nucleares.unam.mx, rafael.olmedo@ciencias.unam.mx}
\vspace{0.2cm}
\end{center}

\begin{center}
{\bf Abstract}
\end{center}
\noindent
In recent years, interesting curved-space extensions of nonrelativistic (NR) string theory have been very actively pursued, where the background has a structure that is a stringy generalization of Newton-Cartan geometry. Here we show that the natural black branes of the NR theory, sourced by the familiar repertoire of stringy objects, generally have a different structure. The black string is our main example. We find that the source distorts the background significantly, generating a large throat within which physics is in fact \emph{relativistic}. It is only far away from the throat that the background approaches the string Newton-Cartan form. We show that exactly the same is true for the longitudinal RR-charged black brane. On the other hand, the transverse RR-charged black brane turns out to have a proper string Newton-Cartan structure everywhere, not just asymptotically.   
 \vspace{0.4cm}
\smallskip
\end{titlepage}

\tableofcontents

\section{Introduction and Summary}\label{introsec}

Around the turn of the century, various  limits of string/M theory were examined, both for their intrinsic value and as potentially enlightening corners of the overall framework. One of the lines of development led to the discovery \cite{Danielsson:2000gi,Gomis:2000bd} of an interesting simplified version of string theory, known as nonrelativistic (NR) string theory because much of the physics in it has that character. In particular, gravity within it, originally thought to be absent, turned out to be present in the form of an instantaneous Newtonian potential \cite{Gomis:2000bd,Danielsson:2000mu}. NR string theory was thus understood to be a UV-complete theory of nonrelativistic quantum gravity, related by an intricate web of dualities \cite{Danielsson:2000gi,Gomis:2000bd,Bergshoeff:2018yvt,Bergshoeff:2022iss,Gomis:2023eav}
to other theories of significant interest, such as the discrete lightcone quantization (DLCQ) of string/M theory \cite{Banks:1996vh,Dijkgraaf:1997vv,Susskind:1997cw,Seiberg:1997ad}, noncommutative Yang-Mills (NCYM) \cite{Sheikh-Jabbari:1997qke,Sheikh-Jabbari:1998aur,Seiberg:1999vs} and decoupled subsectors of the AdS/CFT correspondence \cite{Gomis:2005pg,Sakaguchi:2007ba,Harmark:2014mpa,Harmark:2017rpg,Harmark:2018cdl,Harmark:2020vll}. 
In the past two decades, many aspects of NR string theory have been explored, in a large body of work that was reviewed recently in \cite{Oling:2022fft}.  

The original works \cite{Danielsson:2000gi,Gomis:2000bd} analyzed NR string theory on a flat background, from two complementary perspectives, which we will here refer to as `extrinsic' and `intrinsic'. The extrinsic approach  \cite{Danielsson:2000gi} starts  with the familiar, relativistic string theory, focuses there on the feature of interest, and
then considers the effect of the NR limit. The intrinsic approach \cite{Gomis:2000bd} starts instead from a worldsheet action that defines NR string theory without reference to any limit, and works out the feature of interest from it (the action itself having been derived in \cite{Gomis:2000bd} by taking the NR limit once and for all). Additional observations on both approaches were made in \cite{Danielsson:2000mu}. In either approach, two `longitudinal' spacetime directions $x^A$ ($A=0,1$) are distinguished from the remaining, `transverse' directions $x^{A'}$. The longitudinal spatial direction $x^1$
is special in that it ought to be compact, and all on-shell finite-energy states of the theory carry positive fundamental string winding along it \cite{Klebanov:2000pp,Danielsson:2000gi,Gomis:2000bd}. {}From the extrinsic perspective, this is a consequence of the fact that the string NR limit involves a drastic enlargement of the longitudinal directions with respect to those that are transverse. The intrinsic worldsheet action derived by Gomis and Ooguri in \cite{Gomis:2000bd} assembles the longitudinal embedding fields $X^A$ into lightcone combinations $X$ and $\bar{X}$, whose dynamics (in conformal gauge) is restricted by Lagrange multiplier fields $\lambda,\bar\lambda$ to be respectively purely holomorphic and anti-holomorphic.\footnote{In this paper we use the $X,\bar X, \lambda,\bar\lambda$ notation to conform to recent conventions. When comparing with the original work \cite{Gomis:2000bd}, one must identify $X\leftrightarrow\gamma$,
$\bar X\leftrightarrow\bar\gamma$,
$\lambda\leftrightarrow\beta$ and $\bar\lambda\leftrightarrow\bar\beta$.\label{betafoot}}

Some years later, \cite{Gomis:2005pg,Andringa:2012uz} studied specific examples of curved backgrounds for NR string theory. The more systematic exploration began in earnest in \cite{Harmark:2017rpg,Bergshoeff:2018yvt,Harmark:2018cdl,Gomis:2019zyu,Gallegos:2019icg}, and has led to a very rich set of developments \cite{Oling:2022fft}. 
Much of the impetus for this line of work arose from relatively recent  
advances on the Newton-Cartan (NC) formalism,
which was originally put forth \cite{Cartan:1923zea,Cartan:1924yea} to provide a geometrization of Newtonian gravity analogous to that of general relativity. 

The essential NR distinction between time and space is incorporated in NC geometry by replacing the usual relativistic metric $g_{\mu\nu}$ with a `clock' (einbein) one-form $\tau_{\mu}$, a spatial co-metric $H^{\mu\nu}$, 
and a gauge field $m_{\mu}$ that parametrizes a freedom in the metric-compatible connection, and couples to the mass current of point particles.  The absolute time of Newtonian gravity is obtained when
$d\tau=0$, a requirement that turns out to be equivalent to the absence of torsion. After \cite{Andringa:2010it}
showed that NC gravity can be obtained by gauging the Bargmann algebra, it was realized in \cite{Christensen:2013lma,Christensen:2013rfa,Hartong:2014pma,Bergshoeff:2014uea,Hartong:2015zia,VandenBleeken:2017rij}
that by allowing torsion one can go beyond Newtonian gravity and describe, still within a NR relativistic setting, situations with gravitational time dilation. This torsional Newton-Cartan (TNC) framework has $d\tau\neq 0$, and 
 can be obtained from general relativity in two different variants: a `Type I' formulation that follows from either a null reduction \cite{Julia:1994bs,Hartong:2014pma} or, equivalently, an extremal $c\to\infty$ limit \cite{Jensen:2014wha,Bergshoeff:2015uaa,Bergshoeff:2015sic,VandenBleeken:2017rij}; and a `Type II' formulation that is identified through expansion in powers of $1/c^2$ \cite{Hansen:2019pkl,Hansen:2020pqs}. The latter type enforces a `twistless' torsion constraint, $\tau\wedge d\tau=0$, which allows spacetime to be foliated by equal-time slices.  For a recent review of these developments, see \cite{Hartong:2022lsy,Bergshoeff:2022eog}. 

The investigations of NR string theory on curved backgrounds have been based on stringy generalizations of NC geometry that largely parallel the developments summarized in the preceding paragraph. Such setting was first discussed in \cite{Andringa:2012uz}, and is known as string Newton-Cartan (SNC) geometry. In going from particle (i.e., standard) NC to SNC, a co-metric $H^{\mu\nu}$ 
is still employed for the transverse directions, but the fact that the string has two longitudinal directions rather than one entails the duplication of the other geometric fields: $\tau_{\mu}\to\tau^A_{\mu}$, $m_{\mu}\to m^A_{\mu}$, where $A=0,1$ is now a tangent-space index. String theory on SNC backgrounds was studied in \cite{Andringa:2012uz,Bergshoeff:2018yvt,Bergshoeff:2019pij,Gomis:2019zyu,Yan:2019xsf} with the zero-torsion constraint $D_{[\mu}\tau^A_{\nu]}=0$. The torsional backgrounds (TSNC) have been explored both in the Type I \cite{Harmark:2017rpg,Harmark:2018cdl,Gallegos:2019icg,Harmark:2019upf,Gallegos:2020egk,Bergshoeff:2021bmc,Bidussi:2021ujm} and Type II \cite{Hartong:2021ekg,Hartong:2022dsx} variants. The twistless torsion constraint enforced by the latter type takes the form $d\tau^A=\alpha^A{}_B\wedge\tau^B$ (for arbitrary one-forms $\alpha^A{}_{B}$), and allows for a codimension-two spacetime foliation. A prominent special subtype restricts $\alpha^A{}_A=0$.  
Hamiltonian aspects of TSNC have been studied in \cite{Kluson:2018egd,Kluson:2018grx,Kluson:2019ifd,Kluson:2021qqv,Kluson:2021tub}.
Worldvolume actions for branes have been worked out in \cite{Kluson:2020aoq,Kluson:2020kyp,Ebert:2021mfu,Roychowdhury:2023rcb,Roychowdhury:2023bqi}.

In \cite{Harmark:2019upf},  equivalence was established between two TSNC setups that a priori seemed different: the one derived via null reduction \cite{Harmark:2017rpg,Harmark:2018cdl} and that obtained via an extremal $c\to\infty$ limit \cite{Andringa:2012uz,Bergshoeff:2018yvt,Bergshoeff:2019pij}. In \cite{Bergshoeff:2021bmc}, it was shown that the same cancellation between the longitudinal metric and Kalb-Ramond field that occurred at the worldsheet level in \cite{Klebanov:2000pp,Danielsson:2000gi,Gomis:2000bd} allows the NR
limit to be taken directly for the action and equations of motion of the target-space NS-NS fields, leading to a setup with an emergent local scale invariance, compatible with certain torsion constraints that define a dilatation-invariant string Newton-Cartan (DSNC) geometry. The extension to supergravity was worked out in \cite{Bergshoeff:2021tfn}, and involves torsion constraints that define a so-called self-dual DSNC (DSNC$^{-}$, for short). See also \cite{Bergshoeff:2023ogz,Bergshoeff:2023igy,Bergshoeff:2023fcf,Bergshoeff:2022iyb}. Another important recent advance was a reformulation \cite{Bidussi:2021ujm} of Type I TSNC where the two one-form gauge fields $m^A_{\mu}$ are replaced by a single two-form gauge field $m_{\mu\nu\,}$, which couples to the tension current of strings. This rewriting is less redundant, and allows for greater generality, in that no torsion constraints are imposed at the classical level. In Section~\ref{sncsec} we will briefly review the main formulae of NR string theory on a (T)SNC background, mostly following \cite{Bergshoeff:2018yvt,Gomis:2019zyu,Bergshoeff:2019pij}. Readers familiar with this formalism can skip the section entirely.

The developments summarized in the previous two paragraphs have been achieved utilizing  the complementary approaches of the original works \cite{Danielsson:2000gi,Gomis:2000bd}. More specifically, whereas studies such as \cite{Bergshoeff:2018yvt,Gomis:2019zyu,Yan:2019xsf,Gomis:2020fui,Gomis:2020izd} start from the \emph{intrinsic} worldsheet action of \cite{Gomis:2000bd} for NR string theory on flat spacetime  and generalize it to curved backgrounds by exponentiating in the usual manner the vertex operators associated with massless states, papers such as \cite{Harmark:2017rpg,Harmark:2018cdl,Gallegos:2019icg,Harmark:2019upf,Bergshoeff:2019pij,Bidussi:2021ujm} deduce the form of the NR nonlinear $\sigma$-model with the \emph{extrinsic} approach of \cite{Danielsson:2000gi}, going back to the relativistic $\sigma$-model and subjecting it to the NR limit (or, equivalently, to null reduction).  
Inquiries into the general geometric structure associated with curved-space NR string theory motivate the search for concrete examples of such nontrivial backgrounds. Within the TSNC framework, solutions to the relevant equations of motion have been found in \cite{Bergshoeff:2022pzk}.
A different route is to use once again the extrinsic approach of \cite{Danielsson:2000gi}, starting with the well-known black branes of relativistic string theory and applying the NR limit to deduce the form of the corresponding NR black branes. This is what we aim to do in the present paper. 

Given that NR string theory is characterized by the property that on-shell closed strings must have positive longitudinal winding \cite{Klebanov:2000pp,Danielsson:2000gi,Gomis:2000bd}, the gravitational source that most assuredly ought to survive the NR limit is a bundle of  $N\gg 1$ fundamental strings (F1s) extending purely along the longitudinal spatial direction. We will study this setup as our main example. The resulting relativistic curved geometry is of course the Dabholkar-Harvey black string \cite{Dabholkar:1989jt,Dabholkar:1990yf}. In Section~\ref{stringsec} we will take the NR limit of this solution, to obtain in (\ref{ANRBS}) the black string of NR string theory. 

Our main finding is that, contrary to expectation, this background is \emph{not} a solution of the (T)SNC equations of motion worked out in \cite{Gomis:2019zyu,Gallegos:2019icg,Bergshoeff:2019pij,Yan:2019xsf,Bergshoeff:2021bmc,Bergshoeff:2021tfn}, and it does not even possess a proper SNC structure. We identify that it is in fact a solution of the equations of motion of ordinary ten-dimensional supergravity, with the special property that it is \emph{asymptotically} nonrelativistic. In other words, even after the NR limit, the bundle of strings distorts the background so significantly that it generates a large throat within which physics is, just like in the parent string theory, \emph{relativistic}. It is only far away from the throat that the background approaches the SNC form. To complement this insight, we also show that the only way to reduce the Dabholkar-Harvey black string to a background that is of (T)SNC form everywhere and solves the (T)SNC equations of motion is to take the NR $c\to\infty$ limit while simultaneously scaling down the strength of the source according to $N\propto 1/c^4 \to 0$. This can be done formally, but it is not natural from the string theory perspective, because the number $N$ of fundamental strings is a discrete parameter. 

In Section~\ref{longsec} we show that precisely the same statements apply for a black $p$-brane with RR charge \cite{Horowitz:1991cd} extended along the longitudinal spatial direction. The relevant NR solution, generated by a stack of $K$ D$p$-branes carrying $N$ units of F1 winding (equivalently, of electric flux within the D-branes) is given in (\ref{longDp}). It was worked out in \cite{Gopakumar:2000na} for $p=3$, by taking the NR limit. Shortly thereafter, it was generalized to other values of $p$ in \cite{Harmark:2000wv}. As summarized in \cite{Guijosa:2023qym}, the physical import of this background was originally misinterpreted, and then reappraised in \cite{Danielsson:2000mu}, where it was argued to be the field configuration sourced by longitudinal D-branes in NR string theory. This perspective was vindicated very recently in \cite{Guijosa:2023qym}, by showing through explicit computation that D-branes as described by the intrinsic Gomis-Ooguri formalism \cite{Gomis:2000bd} indeed give rise to a curved background that matches the one obtained in \cite{Gopakumar:2000na,Harmark:2000wv}. As explained in \cite{Danielsson:2000mu}, strings with any winding can exist within the relativistic throat, but only those with positive winding can escape to the asymptotic NR region.  

In Section~\ref{transsec}, we study these same questions for a transverse RR black brane, i.e., one that is not extended along the longitudinal spatial direction. Interestingly, in that case the situation ends up being different: the NR transverse black $p$-brane does have a proper SNC structure, given in (\ref{yessnc}). In other words, transverse D$p$-branes do not generate a $\lambda\bar\lambda$ deformation, a fact that is visible in the recent results of \cite{Guijosa:2023qym}. Ultimately, this atypical, purely SNC character of the transverse black brane is due to the exceptional absence of F1 winding, a feature identified already in \cite{Danielsson:2000gi,Danielsson:2000mu}, and recalled in Section~\ref{transsec}.
Also worth highlighting is the fact that the harmonic function (\ref{deltafcn}) appearing in the transverse black $p$-brane has the peculiar feature of being completely localized, such that the background is flat SNC  everywhere except on the 8-dimensional plane $x^1=0$, within which it is only asymptotically flat. 

Several of the papers on NR string theory of recent years, most notably \cite{Gomis:2019zyu,Gallegos:2019icg,Yan:2019xsf,Blair:2020ops,Gomis:2020fui,Yan:2021lbe,Oling:2022fft}, 
have mentioned the possibility of interpolating between nonrelativistic and relativistic string theory in parameter space, by adding to the Gomis-Ooguri action \cite{Gomis:2000bd} a 
$\lambda\bar\lambda$ term (where $\lambda,\bar\lambda$ are the Lagrange multiplier fields mentioned above, in the main text and in footnote \ref{betafoot}).\footnote{Particularly noteworthy is the fact that this $\lambda\bar\lambda$ deformation has been shown to be directly analogous to the much studied $T\bar T$ deformatión \cite{Blair:2020ops}.} To date, this has been brought up within the view that NR string theory should be \emph{defined} by forbidding the appearance of such a term. This would apply at two distinct levels: aside from switching off the $\lambda\bar\lambda$ coupling classically, one needs to constrain the background appropriately to ensure that the $\lambda\bar\lambda$ coupling does not run quantum-mechanically (i.e., its beta functional must be set to zero). One elegant possibility is to achieve this via a symmetry principle, as was first suggested in \cite{Gomis:2019zyu} and has been further explained and explored in \cite{Yan:2021lbe}.

The perspective of defining NR string theory as a self-contained formalism with TSNC structure, purposefully disconnected from the familiar relativistic setting, has proven to be very fertile. The results of the present paper, however, align with a different perspective. The stringy sources examined here have a rightful claim to be considered part of the spectrum of NR string theory, and we are learning that they call for a wider definition of the formalism, that incorporates backgrounds where the $\lambda\bar\lambda$ deformation is present but switches off asymptotically. As we stress in Section~\ref{longsec}, this feature is particularly explicit in the recent results of \cite{Guijosa:2023qym}, which showed that the simple act of adding a boundary to the Gomis-Ooguri worldsheet automatically generates a $\lambda\bar\lambda$ contribution to the action. It is likewise seen in our findings in Section~\ref{stringsec}.

{}From this standpoint, \emph{NR string theory is an interesting class of vacua of relativistic string theory}, distinguished by the NR features induced by their \emph{asymptotic} behavior. It is in this asymptopia that all the recent advances on TSNC geometry find their natural home. A very diverse set of backgrounds should exist that interpolate between an inner relativistic region and an outer region that is NR but non-flat.   
We see all of this as another step in a trend that started when NCOS theory \cite{Seiberg:2000ms,Gopakumar:2000na}, initially received as an unexpected instance of a non-gravitational string theory, was subsumed within NR string theory \cite{Klebanov:2000pp,Danielsson:2000gi,Gomis:2000bd}, with gravity that first presented itself in Newtonian attire, but is now revealing its Einsteinian soul.      

\section{(Torsional) String Newton-Cartan Geometries and Nonrelativistic String Theory} \label{sncsec}

\subsection{(T)SNC Geometry}
In this subsection we briefly review the description of (T)SNC geometries, closely following \cite{Bergshoeff:2018yvt,Gomis:2019zyu} and highlighting some facts that will be important for our analysis below.  Let $\mathcal{T}_{P}$ be the tangent space at a point $P$ in the $D$-dimensional spacetime manifold $\mathcal{M}$. Within (T)SNC, $\mathcal{T}_{P}$ is decomposed into two directions, indexed by $A=0,1$, which will be longitudinal to the strings, and $D-2$ transverse directions indexed by $A'=2,\ldots,D-1$. Denoting the longitudinal and transverse vielbein fields by $\tau_{\mu}^{A}$ and $E_{\mu}^{A'}$, the inverses $\tau^{\mu}_{A}$ and $E^{\mu}_{A'}$ are defined through the relations
\begin{subequations}
    \begin{align}
    \tau_{\mu}^{A}\tau^{\mu}_{B}&=\delta^{A}_{B}~,\\
    E_{\mu}^{A'}E^{\mu}_{B'}&=\delta^{A'}_{B'}~,\\
    \tau^{\mu}_{A}E_{\mu}^{A'}=\tau_{\mu}^{A}E^{\mu}_{A'}&=0~,\\
    \tau_{\mu}^{A}\tau^{\nu}_{A}+E_{\mu}^{A'}E^{\nu}_{A'}&=\delta_{\mu}^{\nu}~.
    \end{align}
\label{SNC_Vielbein}
\end{subequations}
These vielbeine determine the longitudinal metric and transverse co-metric 
\begin{subequations}
\label{SNC_MetricsETau}
    \begin{align}
    \tau_{\mu\nu}\equiv \tau_{\mu}^{A}\tau_{\nu}^{B}\eta_{AB}~,\\
    H^{\mu\nu}\equiv E^{\mu}_{A'}E^{\nu}_ {B'}\delta^{A'B'}~.
    \end{align}
\end{subequations}
TSNC geometry is defined by these data together with the two longitudinal gauge fields $m^A_{\mu}\,$, which are associated with a non-central extension $Z_{A}$ of the string Galilei algebra.
In particular, a flat spacetime corresponds to $\tau_{\mu}^{A}=\delta_{\mu}^{A}\,$, $E_{\mu}^{A'}=\delta_{\mu}^{A'}$ and $m_{\mu}^{A}=0$.

A few comments are in order. First, one notes that the metric and co-metric in (\ref{SNC_MetricsETau}) are orthogonal to one another: by virtue of (\ref{SNC_Vielbein}), they satisfy $\tau_{\mu\nu}H^{\nu\lambda}=0$. Second, they 
are invariant under the (string) Galilei boosts,
which leave $\tau^A_{\mu}$ invariant but rotate $\delta_{\scriptstyle\Sigma} E^{A'}_{\mu}=\Sigma^{A'}{}_{\!\!\! A}\,\tau^A_{\mu}\,$, implying through (\ref{SNC_Vielbein}) that
$\delta_{\scriptstyle\Sigma}\tau_A^{\mu}=-\Sigma^{A'}{}_{\!\!\! A}E_{A'}^{\mu}\,$, while
$E_{A'}^{\mu}$ is left invariant.
Third,
when considering the inverses for  $\tau_{\mu\nu}$ and $H^{\mu\nu}$, the obvious choices $\tau^{\mu}_{A}\tau^{\nu}_{B}\eta^{AB}$ and $E_{\mu}^{A'}E_{\nu}^{B'}\delta_{A'B'}$ are not invariant under the Galilei boosts.
However, taking into account that $\delta_{\scriptstyle\Sigma}m_{\mu}^{A}=\Sigma^{A}{}_{\!\!\! A'}\,E^{A'}_{\mu}$, one can form the Galilei boost-invariant combinations
\begin{subequations}
\label{SNC_MetricsNH}
    \begin{align}
    H_{\mu\nu}&\equiv E_{\mu}^{A'}E_{\nu}^{B'}\delta_{A'B'}+\left(\tau_{\mu}^{A}m_{\nu}^{B}+\tau_{\nu}^{A}m_{\mu}^{B}\right)\eta_{AB}~,\\
    \tau^{\mu\nu}&\equiv\tau^{\mu}_{A}\tau^{\nu}_{B}\eta^{AB}-\left(E^{\mu}_{A'}\tau^{\nu}_{A}+E^{\nu}_{A'}\tau^{\mu}_{A}\right)m_{\lambda}^{A}E^{\lambda}_{A'}~.
    \end{align}
\end{subequations}

SNC geometry is characterized by the zero-torsion constraint 
\begin{equation}
D_{[\mu}\tau_{\nu]}^{A}=0 \quad \Longleftrightarrow \quad \partial_{[\mu}\tau_{\nu]}^{A}=\omega_{[\mu}^{AB}\tau_{\nu]}{}_{B}~,
\label{ZeroTorsionV1}
\end{equation}
where $D_{\mu}$ is the covariant derivative with respect to the longitudinal Lorentz boost transformations, and $\omega_{\mu}^{AB}$ is the corresponding spin connection. There are $D$ components of (\ref{ZeroTorsionV1}) that allow one to solve for $\omega_{\mu}^{AB}$ in terms of $\tau_{\mu}^{A}$, while the remaining $D(D-2)$ equations can be written in the form
\begin{equation}
\epsilon_{C}{}^{\left(A \right.}\tau_{\left[\mu\right.}^{\left.B\right)}\partial_{\nu}\tau_{\left.\rho \right]}^{C}=0~.
\label{ZeroTorsionV2}
\end{equation}
As explained in Section~\ref{introsec}, less restrictive torsion constraints, or even fully unconstrained TSNC, have been considered in works such as  \cite{Harmark:2017rpg,Bergshoeff:2021bmc,Yan:2021lbe,Bidussi:2021ujm,Hartong:2022dsx}. For our purposes in the present paper, it will be enough to focus on the SNC case. As we will see in the next section, our findings do not hinge on choosing a particular flavor of TSNC, but refer instead to the much more fundamental contrast between the relativistic and nonrelativistic geometric structures. 

With the fields defined above, it is possible to compute the curvature tensors of SNC geometry. The Christoffel symbols are defined as\footnote{This definition assumes a special choice of the field $W_{ABA'}$ introduced in \cite{Bergshoeff:2019pij}. However, said field drops out of the equations of motion.}
\begin{equation}
\Gamma^{\rho}{}_{\mu\nu}=\frac{1}{2}\tau^{\rho\sigma}(\partial_{\mu}\tau_{\sigma\nu}+\partial_{\nu}\tau_{\sigma\mu}-\partial_{\sigma}\tau_{\mu\nu})
+\frac{1}{2}H^{\rho\sigma}(\partial_{\mu}H_{\sigma\nu}+\partial_{\nu}H_{\sigma\mu}-\partial_{\sigma}H_{\mu\nu})
~.
\label{SNC_Christoffel}
\end{equation}
This connection is metric compatible in the sense that $\nabla_{\mu}\tau_{\nu\rho}=0$, $\nabla_{\mu}H^{\nu\rho}=0$.
{}From (\ref{SNC_Christoffel}) we can compute the Riemann and Ricci tensors
\begin{equation}
R^{\sigma}{}_{\rho\mu\nu}=\partial_{\mu}\Gamma^{\sigma}{}_{\rho\nu}+\Gamma^{\sigma}{}_ {\mu\lambda}\Gamma^{\lambda}{}_{\rho\nu}-\mu\leftrightarrow\nu~, \qquad R_{\mu\nu}=R^{\sigma}{}_{\mu\sigma\nu}~.
\label{SNC_Riemann}
\end{equation}

\subsection{SNC as a NR limit}
The way to obtain SNC geometry as a nonrelativistic limit of General Relativity (GR) was first shown in \cite{Bergshoeff:2019pij}. In what follows, variables with a hat correspond to the relativistic versions, and unhatted ones, to their NR counterparts. Starting from the vielbein formulation of GR, the spacetime metric $\hat{G}_{\mu\nu}$ is written as
\begin{equation}
\hat{G}_{\mu\nu}=\hat{E}_{\mu}^{\hat{A}}\hat{E}_{\nu}^{\hat{B}}\eta_{\hat{A}\hat{B}}~,
\label{GRVielbein}
\end{equation}
where 
\begin{equation}
\hat{E}_{\mu}^{\hat{A}}\hat{E}^{\mu}_{\hat{B}}=\delta^{\hat{A}}_{\hat{B}}~, \quad \hat{E}_{\mu}^{\hat{A}}\hat{E}^{\nu}_{\hat{A}}=\delta^{\nu}_{\mu}~.
\label{RelativisticVielbein}
\end{equation}
In order to properly define the NR limit, one needs to add an auxiliary two-form gauge field $\hat{M}_{\mu\nu}$ with zero curvature. In the string theory setting, this will be the near-critical Kalb-Ramond field that plays an important role in the definition of NR string theory, as seen in (\ref{BM}) below.  Introducing the speed of light $c$, one then postulates that the metric and two-form field scale as follows: 
\begin{subequations}
\label{NR_VielbeinLimit}
    \begin{align}
    \hat{E}_{\mu}^{\hat{A}}&=c\tau_{\mu}^{A}+\frac{1}{c}\left(m_{\mu}^{A}-C_{\mu}^{A}\right)~,
    \\
\hat{E}_{\mu}^{\hat{A'}}&=E_{\mu}^{A'}~,
\\
    \hat{M}_{\mu\nu}&=-\left(c\tau_{\mu}^{A}-\frac{1}{c}C_{\mu}^{A}\right)\left(c\tau_{\nu}^{B}-\frac{1}{c}C_{\nu}^{B}\right)\epsilon_{AB}~.
    \end{align}
\end{subequations}
Here $A=0,1$ and $A'=2,\ldots,D-1$ as before, and $\tau_{\mu}^{A}$, $E_{\mu}^{A'}$ and $m_{\mu}^{A}$ are the SNC fields introduced in the previous subsection. The $C_{\mu}^{A}$ are arbitrary functions necessary to correctly take the NR limit, but they drop out of the equations of motion, which we will review in the next subsection. (This redundancy is eliminated in the new formalism proposed recently in \cite{Bidussi:2021ujm}.) Note that applying the $c\rightarrow\infty$ limit to (\ref{RelativisticVielbein}) consistently leads to (\ref{SNC_Vielbein}). The expansions (\ref{NR_VielbeinLimit}) together with (\ref{GRVielbein}) imply that the large-$c$ expansion of $\hat{G}_{\mu\nu}$ is
\begin{equation}
\hat{G}_{\mu\nu}=c^{2}\tau_{\mu\nu}+\left[H_{\mu\nu}-\left(\tau_{\mu}^{A}C_{\nu}^{B}+\tau_{\nu}^{A}C_{\mu}^{B}\right)\eta_{AB}\right]+\frac{1}{c^{2}}\left(m_{\mu}^{A}-C_{\mu}^{A}\right)\left(m_{\nu}^{B}-C_{\nu}^{B}\right)\eta_{AB}~.
\label{NRExpansion_G}
\end{equation}
Direct computation of the Christoffel symbols and the Riemann tensor associated with the metric (\ref{NRExpansion_G}) leads to
\begin{equation}
\hat{\Gamma}^{\rho}{}_{\mu\nu}=\Gamma^{\rho}{}_{\mu\nu}+\mathcal{O}\left(\frac{1}{c^{2}}\right), \quad \hat{R}^{\sigma}{}_{\rho\mu\nu}=R^{\sigma}{}_{\rho\mu\nu}+\mathcal{O}\left(\frac{1}{c^{2}}\right),
\end{equation}
where $\Gamma^{\rho}{}_{\mu\nu}$ and $R^{\sigma}{}_{\rho\mu\nu}$ are precisely the Christoffel symbols and Riemann tensor of the SNC geometry defined in (\ref{SNC_Christoffel}) and (\ref{SNC_Riemann}).

\subsection{Strings on SNC backgrounds: intrinsic approach}

We are interested in string propagation on backgrounds that involve a SNC geometry coupled to the Kalb-Ramond and dilaton fields. Alluding to the intrinsic vs.~extrinsic terminology defined in Section~\ref{introsec}, in this subsection we briefly review the $\sigma$-model action and associated equations of motion obtained through the intrinsic approach of \cite{Gomis:2019zyu,Bergshoeff:2019pij}.

NR string theory is intrinsically defined by the Gomis-Ooguri action \cite{Gomis:2000bd} on a Riemann surface $\Sigma$, the worldsheet of the string, which we parameterize by $\sigma^{\alpha}$, with $\alpha=1,2$. We denote the worldsheet metric by $h_{\alpha\beta}$, and introduce the zweibein $e_{\alpha}^{a}$, with $a=1,2$, such that $h_{\alpha\beta}=e_{\alpha}^{a}e_{\beta}^{b}\delta_{ab}$. On the worldsheet one has the string embedding fields $X^{\mu}(\sigma)$ and the two auxiliary fields $\lambda(\sigma)$ and $\bar{\lambda}(\sigma)$ that are essential to the Gomis-Ooguri formalism.

To describe a nontrivial background involving a curved spacetime geometry $\tau^A_{\mu}$, $H_{\mu\nu}$ together with Kalb-Ramond and dilaton fields $B_{\mu\nu}$ and $\Phi$, one must exponentiate
in the usual fashion the vertex operators associated with the massless modes of the string. The resulting path integral takes the form\footnote{Left implicit here are the ghosts from gauge-fixing, as well as the worldsheet superpartners \cite{Gomis:2000bd,Gomis:2004pw,Gomis:2005pg,Kim:2007hb,Kim:2007pc,Park:2016sbw,Blair:2019qwi}.\label{implicitfieldsfoot}}
\begin{equation}
Z=\int\mathscr{D}\lambda\mathscr{D}\bar{\lambda}\mathscr{D}X^{\mu} \exp\left[-S_{\sigma}\right]~,
\label{NRpathintegral}
\end{equation}
with the $\sigma$-model action
\cite{Gomis:2019zyu,Bergshoeff:2018yvt,Bergshoeff:2019pij}
\begin{equation}
\begin{aligned}
S_{\sigma}=&\frac{1}{4\pi\alpha'}\int d^{2}\sigma\sqrt{h}\left[\lambda\bar{\mathcal{D}}X^{\mu}\tau_{\mu}+\bar{\lambda}\mathcal{D}X^{\mu}\bar{\tau}_{\mu}+\mathcal{D}X^{\mu}\bar{\mathcal{D}}X^{\nu}\left(H_{\mu\nu}+B_{\mu\nu}\right)\right]\\
&+\frac{1}{4\pi}\int d^{2}\sigma\sqrt{h} R^{(2)}\left(\Phi-\frac{1}{4}\ln G\right)~.
\end{aligned}
\label{NRSigmaAction}
\end{equation}
Here, $\alpha'$ is the Regge slope of the NR string theory, $h=\text{det}(h_{\alpha\beta})$, $h^{\alpha\beta}$ is the inverse worldsheet metric, $R^{(2)}$ is the Ricci scalar of $h_{\alpha\beta}\,$,
\begin{equation}
\tau_{\mu}\equiv \tau_{\mu}^{0}+\tau_{\mu}^{1}~, \qquad \bar{\tau}_{\mu}\equiv \tau_{\mu}^{0}-\tau_{\mu}^{1}~,
\end{equation}
and
\begin{equation}
G\equiv \text{det}^{(D-2)}\left(H_{\mu\nu}\right)\text{det}^{(2)}\left(\tau_{\rho}^{A}H^{\rho\sigma}\tau_{\sigma}^{B}\right)~.
\label{detg}
\end{equation}
Also employed in (\ref{NRSigmaAction}) are the differential operators
\begin{equation}
\mathcal{D}\equiv \frac{i}{\sqrt{h}}\epsilon^{\alpha\beta}\bar{e}_{\alpha}\nabla_{\beta}~, \qquad \bar{\mathcal{D}}\equiv \frac{i}{\sqrt{h}}\epsilon^{\alpha\beta}e_{\alpha}\nabla_{\beta}~,
\end{equation}
where $\nabla_{\alpha}$ is the covariant derivative compatible with $h_{\alpha\beta}$, $\epsilon^{\alpha\beta}$ is the two-dimensional Levi-Civita symbol, and
\begin{equation}
e_{\alpha}\equiv e_{\alpha}^{1}+ie_{\alpha}^{2}, 
\qquad 
\bar{e}_{\alpha}\equiv-e_{\alpha}^{1}+ie_{\alpha}^{2}~.
\end{equation}
At this stage, it is important to note that the action (\ref{NRSigmaAction}) is invariant under the $Z_{A}$ transformations only if the zero torsion constraint (\ref{ZeroTorsionV1}) is satisfied. This $Z_{A}$ symmetry forbids the $\lambda\bar{\lambda}$ operator from being generated in the world sheet action. As mentioned in Section~\ref{introsec}, this operator has the effect of deforming nonrelativistic string theory towards relativistic string theory.

As usual, the background must be constrained by requiring that the path integral (\ref{NRpathintegral}) be Weyl invariant, in order for string propagation to be consistently defined. This amounts to the vanishing of the beta functionals for the couplings seen in (\ref{NRSigmaAction}). Explicitly, one must have \cite{Gomis:2019zyu}
\begin{subequations}
\label{SNC_EOM}
    \begin{align}
    \label{SNC_EOM1}
    & P_{A'B'}=Q_{A'B'}=P_{A A'}+\epsilon_{A}{}^{B}Q_{BA'}=\eta^{AB}P_{AB}-\epsilon^{AB}Q_{AB}=0~,
    \\
    \label{SNC_DilatonEOM}
    & \nabla_{A'}\nabla^{A'}\Phi-\nabla^{A'}\Phi\nabla_{A'}\Phi+\frac{1}{4}R_{A'}{}^{A'}-\frac{1}{48}\mathcal{H}_{A'B'C'}\mathcal{H}^{A'B'C'}=0~,
    \end{align}
\end{subequations}
where 
\begin{subequations}
\label{PQ_Def}
    \begin{align}
    \label{calH_Def}
    \mathcal{H}_{\mu\nu\rho}&\equiv \partial_{\mu}B_{\nu\rho}+\partial_{\rho}B_{\mu\nu}+\partial_{\nu}B_{\rho\mu}~,
    \\
    \label{P_Def}
    P_{\mu\nu}&\equiv R_{\mu\nu}+2\nabla_{\mu}\nabla_ {\nu}\Phi-\frac{1}{4}\mathcal{H}_{\mu A' B'}\mathcal{H}_{\nu}{}^{A'B'}~,
    \\
    \label{Q_Def}
    Q_{\mu\nu}&\equiv -\frac{1}{2}\nabla^{A'}\mathcal{H}_{A'\mu\nu}+\mathcal{H}_{A'\mu\nu}\nabla^{A'}\Phi~.
    \end{align}
\end{subequations}
These are the equations of motion for SNC gravity coupled to the Kalb-Ramond and dilaton fields, at one-loop order in the $\alpha'$ expansion. They must hold together with the zero-torsion constraint (\ref{ZeroTorsionV2}), which arises by demanding that the beta functions of $\tau_{\mu}^{A}$ vanish. In (\ref{SNC_EOM}), anytime a spacetime index is contracted with the spacetime index of a vielbein with no derivatives acting on it, said index has been replaced with the flat index of this vielbein field. For instance,
\begin{equation}
P_{A'B'}=E^{\sigma}_{A'}E^{\rho}_{B'}P_{\sigma\rho}~, \quad Q_{A'B'}=E^{\sigma}_{A'}E^{\rho}_{B'}Q_{\sigma\rho}~.
\end{equation}

\subsection{Strings on SNC backgrounds: extrinsic approach}

Another way to obtain the SNC equations of motion is by starting with relativistic string theory and following the effect of the limit. This can be done at two levels, leading to the same result. The first option is to analyze the NR limit of the relativistic $\sigma$-model worldsheet action, and show that this leads to (\ref{NRSigmaAction}). The other option is to obtain the beta functions of relativistic string theory and then take the NR limit at this stage. This of course should lead to the equations of motion listed in (\ref{SNC_EOM}). 

We will briefly review here the NR limit of the worldsheet action, as some aspects will be contrasted with our results below. The starting point is the path integral
\begin{equation}
\hat{Z}=\int \mathscr{D} X^{\mu}\exp\left[-\hat{S}_{\sigma}\right]~,
\end{equation}
where the relativistic $\sigma$-model action in the presence of the spacetime string-frame metric $\hat{G}_{\mu\nu}$, Kalb-Ramond field $\hat{B}_{\mu\nu}$ and dilaton field $\hat{\Phi}$ takes the familiar form
\begin{equation}
\begin{aligned}
\hat{S}_{\sigma}=&\frac{1}{4\pi\alpha'}\int d^{2}\sigma\,\partial_{\alpha}X^{\mu}\partial_{\beta}X^{\nu}\left(\sqrt{h}h^{\alpha\beta}\hat{G}_{\mu\nu}+i\epsilon^{\alpha\beta}\hat{B}_{\mu\nu}\right)\\&+\frac{1}{4\pi}\int d^{2}\sigma\sqrt{h} R^{(2)}\left(\hat{\Phi}-\frac{1}{4}\ln (-\hat{G})\right)
\\
=&\frac{1}{4\pi\alpha'}\int d^{2}\sigma\sqrt{h}\left[\mathcal{D}X^{\mu}\bar{\mathcal{D}}X^{\nu}\left(\hat{G}_{\mu\nu}+\hat{B}_{\mu\nu}\right)\right]\\
&+\frac{1}{4\pi}\int d^{2}\sigma\sqrt{h} R^{(2)}\left(\Phi-\frac{1}{4}\ln (-G)\right)~.\end{aligned}
\label{relativisticsigma}
\end{equation}
with $\hat{G}=\text{det}\left(\hat{G}_{\mu\nu}\right)$. 

One then considers the following NR expansions for the fields,
\begin{equation}
\hat{B}_{\mu\nu}=\hat{M}_{\mu\nu}+B_{\mu\nu}~, \quad \hat{\Phi}=\ln c+\Phi~,
\label{BM}
\end{equation}
which, together with the expansions given in (\ref{NR_VielbeinLimit}) and in (\ref{NRExpansion_G}) (we list the latter again here for ease of consultation), explicitly leads to \cite{Bergshoeff:2019pij}
\begin{align}
\hat{G}_{\mu\nu}&=c^{2}\tau_{\mu\nu}+
\left[
H_{\mu\nu}-\left(\tau_{\mu}^{A}C_{\nu}^{B}+\tau_{\nu}^{A}C_{\mu}^{B}\right)\eta_{AB}
\right]
+\frac{1}{c^{2}}\left(m_{\mu}^{A}-C_{\mu}^{A}\right)\left(m_{\nu}^{B}-C_{\nu}^{B}\right)\eta_{AB}~,
\nonumber\\
\hat{B}_{\mu\nu}&=-c^{2}\tau_{\mu}^{A}\tau_{\nu}^{B}\epsilon_{AB}+\left[B_{\mu\nu}+\left(\tau_{\mu}^{A}C_{\nu}^{B}-\tau_{\nu}^{A}C_{\mu}^{B}\right)\epsilon_{AB}\right]-\frac{1}{c^{2}}C_{\mu}^{A}C_{\nu}^{B}\epsilon_{AB}~,
\label{NRExpansion_GBPHi}\\ 
\hat{\Phi}&=\ln c+\Phi~.
\nonumber
\end{align}
Using these expansions, it is possible to show \cite{Bergshoeff:2019pij} that the action can be reexpressed as
\begin{align}
\hat{S}_{\sigma}=&\frac{1}{4\pi\alpha'}\int d^{2}\sigma\sqrt{h}\Biggr\{\Biggr[\lambda\bar{\mathcal{D}}X^{\mu}\tau_{\mu}+\bar{\lambda}\mathcal{D}X^{\mu}\bar{\tau}_{\mu}
+\mathcal{D}X^{\mu}\bar{\mathcal{D}}X^{\nu}\left(H_{\mu\nu}+B_{\mu\nu}\right)
\nonumber\\
&\qquad\qquad\qquad\qquad
+\left.\alpha'R^{(2)}\left(\Phi-\frac{1}{4}\ln\left(-\frac{\hat{G}}{c^{4}}\right)\right)\right]
\label{nonrelativisticsigma}\\
&\qquad\qquad\quad\;
+\frac{1}{c^{2}}\Biggl[\left(\lambda\bar{\lambda}-\lambda\bar{\mathcal{D}}X^{\mu}C_{\mu}-\bar{\lambda}\mathcal{D}X^{\mu}\bar{C}_{\mu}\right)
\nonumber\\
&\qquad\qquad\qquad\qquad
+\mathcal{D}X^{\mu}\bar{\mathcal{D}}X^{\nu}\left(m_{\mu}^{A}m_{\nu}^{B}-m_{\mu}^{A}C_{\nu}^{B}-C_{\mu}^{A}m_{\nu}^{B}\right)\eta_{AB}\Bigg]\Biggr\}~,
\nonumber
\end{align}
where we have defined
\begin{equation}
C_{\mu}\equiv C_{\mu}^{0}+C_{\mu}^{1}~, 
\quad 
\bar{C}_{\mu}\equiv C_{\mu}^{0}-C_{\mu}^{1}~.
\label{NRExpansion_SG}
\end{equation}

As exhibited in \cite{Bergshoeff:2019pij},
(\ref{NRExpansion_G}) and (\ref{detg}) imply that
\begin{equation}
 -\lim_{c\to\infty}\frac{\hat{G}}{c^4}=\text{det}^{(D-2)}\left(H_{\mu\nu}\right)\text{det}^{(2)}\left(\tau_{\rho}^{A}H^{\rho\sigma}\tau_{\sigma}^{B}\right)\equiv G~,   
\end{equation}
and using this result, the NR limit of (\ref{NRExpansion_SG}) is precisely (\ref{NRSigmaAction}). Note that the auxiliary fields $C_{\mu}^{A}$ drop out from the action in this limit, together with the $\lambda\bar{\lambda}$ term. This confirms that when the $\lambda\bar{\lambda}$ term is present we are indeed in relativistic string theory.

\section{NR Black String} 
\label{stringsec}
Having reviewed in the previous section
the description of strings propagating on (T)SNC backgrounds, we now proceed to the main objective of this paper, which is to work out some explicit curved backgrounds belonging to NR string theory. As explained in Section~\ref{introsec}, we will opt for an extrinsic approach, selecting well-known black brane solutions in relativistic string theory and then following the effect of the NR limit. Since positive longitudinal fundamental string (F1) winding is required to survive the limit \cite{Klebanov:2000pp,Danielsson:2000gi,Gomis:2000bd}, the main example we choose to examine, which will be the topic of the present section, is the static extremal black string \cite{Dabholkar:1989jt,Dabholkar:1990yf} extended along $\hat{x}^1\equiv\hat{x}$. As before, hatted symbols refer to quantities before the limit.    

In the string frame, 
the black string is described by the Dabholkar-Harvey (DH) metric, Kalb-Ramond two-form and dilaton
\begin{subequations}
\begin{align}
d\hat{s}^{2}&=\mathsf{h}^{-1}\left(-d\hat{t}^{2}+d\hat{x}^{2}\right)
+d\hat{x}_2^2+\ldots+d\hat{x}_9^2~,
\label{BS_Metric}\\
\hat{B}&=\mathsf{h}^{-1}d\hat{t}\wedge d\hat{x}~,
\label{BS_B} \\
\exp\left(2\hat{\Phi}\right)&=\hat{g}_{s}^{2}\,\mathsf{h}^{-1}~,
\label{BS_Phi}
\end{align}
\label{BS_full}
\end{subequations}
where
\begin{subequations}
\begin{align}
\mathsf{h}&=1+\frac{\hat{L}^{6}}{\hat{r}^{6}}~, 
\\
\hat{r}^2&\equiv \hat{x}_2^2+\ldots+\hat{x}_9^2~,
\\
\hat{L}^{6}&\equiv 32\pi^{2}\hat{N} \hat{g}_{s}^{2}\hat{l}_{s}^{\,6}~,
\end{align}
\label{BS_harmonic}
\end{subequations}
\kern-0.35em  $(\hat{t},\hat{x})$ are the directions longitudinal to the string, and $(\hat{x}_2,\ldots,\hat{x}_{9})$ denote the transverse directions. The curvature radius $L$ is determined by the number $\hat{N}\in\mathbb{N}$ of units of F1 charge, the relativistic string coupling $\hat{g}_{s}$, and the string length $\hat{l}_{s}$. 
Expressions (\ref{BS_full}) constitute a solution to the equations of motion of the NS-NS sector of ten-dimensional supergravity, 
\begin{subequations}
\begin{align}
\hat{R}_{\mu\nu}+2\nabla_{\mu}\nabla_{\nu}\hat{\Phi}-\frac{1}{4}\hat{\mathcal{H}}_{\mu\alpha\beta}\hat{\mathcal{H}}_{\nu}{}^{\alpha\beta}&=0~,
\\
-\nabla^{\mu}\nabla_{\mu}\hat{\Phi}+\nabla^{\mu}\hat{\Phi}\nabla_{\mu}\hat{\Phi}-\frac{1}{4}\hat{R}+\frac{1}{48}\hat{\mathcal{H}}_{\mu\nu\alpha}\hat{\mathcal{H}}^{\mu\nu\alpha}&=0~,
\\
-\frac{1}{2}\nabla^{\alpha}\hat{\mathcal{H}}_{\alpha\mu\nu}+\hat{\mathcal{H}}_{\alpha\mu\nu}\nabla^{\alpha}\hat{\Phi}&=0~.
\end{align}
\label{NSNSSUGRAEOM}
\end{subequations}

The string NR limit is $c\to\infty$ with the following scaling for the parameters and coordinates:\footnote{For historical reasons, the original works \cite{Danielsson:2000gi,Gomis:2000bd,Danielsson:2000mu} phrased the NR limit as $\delta\to 0$ in units where the relativistic string length $\hat{l}_s=\sqrt{\delta}\,l_s\to 0$. To facilitate comparison with the recent literature, we choose instead to follow here the convention of working in units where the string length is held constant. To translate to the old convention, we must identify $\delta\equiv 1/c^2$ and equate dimensionless ratios. The scaling (\ref{NRBS_Scaling}) then becomes $\hat{l}_s=l_s/c$, $\hat{g}_s=c\,g_s$, $\hat{t}=t$, $\hat{x}=x$, $\hat{r}=r/c$, $\hat{\Omega}_7=\Omega_7$. In either convention, the physical effect is to enlarge the longitudinal directions with respect to the transverse ones, and the string worldsheet action takes the same overall value, on account of being dimensionless. \label{translationfoot}} 
\begin{equation}
\hat{l}_s=l_s~,\quad 
\hat{g}_{s}=c\,g_{s}~, \quad
\hat{N}=N~, \quad
\hat{t}=c\,t~,\, 
\hat{x}=c\,x~, \quad 
\hat{x}_2=x_2~, \ldots,\,
\hat{x}_{9}=x_9~.
\label{NRBS_Scaling}
\end{equation}
The integer $\hat{N}$ is naturally held fixed. 
This implies that $\hat{r}=r$ and $\hat{L}^6=L^{6}c^{2}$, with 
\begin{equation}
L^6\equiv 32\pi^{2}N g_{s}^{2}l^{\,6}_{s} 
\label{L}
\end{equation}independent of $c$. Applying (\ref{NRBS_Scaling}) to (\ref{BS_full}), we obtain the expansion 
\begin{subequations}
\label{ANRBS_Expansion}
\begin{align}
d\hat{s}^{2}&=\left(\frac{r^{6}}{L^{6}}-\frac{r^{12}}{L^{12}c^{2}}+\mathcal{O}\left(\frac{1}{c^{4}}\right)\right)\left(-dt^{2}+dx^{2}\right)+dx_2^{2}+\ldots +dx_9^2~,
\\
\hat{B}&=\left(\frac{r^{6}}{L^{6}}-\frac{r^{12}}{L^{12}c^{2}}+\mathcal{O}\left(\frac{1}{c^{4}}\right)\right)dt\wedge dx~,\\
\hat{\Phi}&=\frac{1}{2}\ln\left(\frac{g_{s}^{2}r^{6}}{L^{6}}\right)-\frac{r^{6}}{2L^{6}c^{2}}+\mathcal{O}\left(\frac{1}{c^{4}}\right)~.
\end{align}
\end{subequations}
We see here that all leading terms are finite, so we can directly identify the line element, Kalb-Ramond two-form and dilaton with their unhatted counterparts. Taking the limit $c\rightarrow\infty$ exactly, we arrive at the final expressions
\begin{subequations}
\begin{align}
ds^{2}&=\frac{r^{6}}{L^{6}}\left(-dt^{2}+dx^{2}\right)+dx_2^{2}+\ldots+dx_9^2~,
\\
B&=\frac{r^{6}}{L^{6}}\,dt\wedge dx~,\\
\exp\left(2\Phi\right)&=\frac{g_{s}^{2}r^{6}}{L^{6}}~.
\end{align}
\label{ANRBS}
\end{subequations}
\emph{By construction, this background describes a black string in NR string theory.} 

 If we try to read off the SNC fields from the scaling seen in (\ref{ANRBS_Expansion}), contact with $\hat{B}$ and with the $\cO(c^2)$ piece in (\ref{NRExpansion_GBPHi}) would lead to identifying 
$E_{\mu}^{A'}=\delta_{\mu}^{A'}$,
$B=\frac{r^{6}}{L^{6}}\,dt\wedge dx$ and $ \tau_{\mu}^{A}=0$. On top of this degeneracy, there is no way to get a match with $\hat{G}_{\mu\nu}$ at $\cO(c^0)$, and $\hat{\Phi}$ is missing the leading $\ln c$ divergence. 
So the background (\ref{ANRBS}) is \emph{not} a solution to the SNC equations of motion (\ref{SNC_EOM}), or to some TSNC modification thereof.

Given that the black string (\ref{ANRBS}) does not have a proper SNC structure, in what sense does it belong to the spectrum of NR string theory? By inspection, we see that the metric is such that the characteristic NR enlargement, as in (\ref{NRBS_Scaling}), of the longitudinal directions with respect to those that are transverse occurs at $r\gg L$,   
 suggesting that (\ref{ANRBS}) asymptotes to a SNC background. In order to verify this, we define
\begin{equation}
\mathbf{c}^{2}\equiv \frac{r^{6}}{L^{6}}~,
\end{equation}
so that $r\gg L$ corresponds to $\mathbf{c}\gg 1$. Through this definition,  (\ref{ANRBS}) is directly rewritten as
\begin{subequations}
\begin{align}
& d\hat{s}^{2}=\mathbf{c}^{2}\left(-dt^{2}+dx^{2}\right)+dx_2^{2}+\ldots+dx_9^2~,
\\
& \hat{B}=\mathbf{c}^{2}\,dt\wedge dx~,
\\
& \hat{\Phi}=\ln \mathbf{c}+\ln g_{s}~,
\end{align}
\label{BSasymptotic}
\end{subequations}
\kern-0.35em where we have restored for a moment the optional hats in the left-hand sides. 
Comparison with  (\ref{NRExpansion_GBPHi}) shows that the $\mathbf{c}\to\infty$ limit yields the flat SNC geometry
\begin{equation}
\tau_{\mu}^{A}=\delta_{\mu}^{A}~, \quad E_{\mu}^{A'}=\delta_{\mu}^{A'}~, \quad m_{\mu}^{A}=C_{\mu}^{A}=0~, \quad 
B_{\mu\nu}=0~, \quad \Phi=\ln g_{s}~.
\label{flatsnc}
\end{equation} 

Moreover, since (\ref{BSasymptotic}) in the $\mathbf{c}\to\infty$ limit is precisely the background that defines NR string theory on flat spacetime \cite{Danielsson:2000gi,Gomis:2000bd},\footnote{See footnote \ref{translationfoot}.}  we know that the relativistic $\sigma$-model (\ref{relativisticsigma}) can be rewritten as in \cite{Gomis:2000bd}
 to yield the NR $\sigma$-model (\ref{nonrelativisticsigma}) on the flat SNC background (\ref{flatsnc}), with $\mathbf{c}$ playing the role of $c\,$. This teaches us two important lessons. One is that, on the background (\ref{ANRBS}), only positively-wound strings can escape to $r\to\infty$ ($\mathbf{c}\to\infty$), and such strings sense a flat SNC geometry, even though (\ref{ANRBS}) has lost the asymptotically flat region (the 1 in the harmonic function $\mathsf{h}(r)$) as a result of the limit (\ref{NRBS_Scaling}).
 The second, equally important lesson is that at $r\, {\scriptstyle\lesssim}\, L$ ($\mathbf{c}\, {\scriptstyle\lesssim}\, 1$) the $1/\mathbf{c}^2$ terms in
 (\ref{nonrelativisticsigma})
 cannot be neglected, and these include the $\lambda\bar\lambda$ term that, as we have mentioned several times, is known to drive the theory to the relativistic regime. In consonance with this, we find by direct calculation that the background (\ref{ANRBS}) is in fact a solution to the full \emph{relativistic} NS-NS supergravity equations of motion (\ref{NSNSSUGRAEOM}).
 
 We conclude then that the NR black string (\ref{ANRBS}) is globally relativistic, but \emph{asymptotically} nonrelativistic. In other words, it is NR due to boundary conditions: the limit (\ref{NRBS_Scaling}) converts the asymptotically Minkowski region to asymptotically string Newton-Cartan. It is this external region that determines the spectrum of asymptotic states of the theory. But the bundle of $N$ strings that generate the relativistic background (\ref{BS_full}) remains a strong source for the NS-NS fields even after the limit, and as a result, (\ref{ANRBS}) contains a large throat inside of which the physics is that of ordinary, relativistic string theory. In particular, there is no constraint on string winding in that region. 

Having learned all of this, we should ask whether there is a way to  modify the scaling (\ref{NRBS_Scaling}) to attenuate the strength of our NS-NS source, in order to ensure that the relativistic black string (\ref{BS_full}) reduces to a background that is SNC \emph{everywhere}, not just asymptotically. To achieve a $1/c^2$ expansion precisely as in   (\ref{NRExpansion_GBPHi}), and knowing that in (\ref{BS_full})  the only remaining parameter at our disposal is $\hat{N}$, we could choose to define the string NR limit as
\begin{equation}
\hat{l}_s=l_s~,\quad 
\hat{g}_{s}=c\,g_{s}~, \quad
\hat{N}=c^{-4}N~, \quad
\hat{t}=c\,t~,\, 
\hat{x}=c\,x~, \quad 
\hat{x}_2=x_2~, \ldots,\,
\hat{x}_{9}=x_9~.
\label{NRBS_Scaling2}
\end{equation}
Through (\ref{NRExpansion_GBPHi}) and (\ref{L}) this entails 
\begin{equation}
\begin{aligned}
& \tau_{\mu}^{A}=\delta_{\mu}^{A}~, \quad 
E_{\mu}^{A'}=\delta_{\mu}^{A'}~, \quad
C_{\mu}^{A}=\frac{L^{6}}{r^{6}}\delta_{\mu}^{A}~, \quad 
m_{\mu}^{A}=\frac{1}{2}C_{\mu}^{A}~, 
\\
& B=\frac{L^{6}}{r^{6}}\,dt\wedge dx~, \quad 
\Phi=\ln g_{s}~,
\end{aligned}
\end{equation}
which is indeed a solution to (\ref{SNC_EOM}). Within the classical SNC perspective, this background is perfectly sensible in itself. It in fact coincides with (a particular case of) the half-BPS `unwinding string' solution of DSNC$^{-}$ derived in \cite{Bergshoeff:2022pzk}, which happens to be torsionless. {}From our extrinsic perspective, however, the usual charge quantization condition \cite{Nepomechie:1984wu,Teitelboim:1985yc} in the parent relativistic theory requires that the number $\hat{N}$ of units of fundamental string charge be an integer, implying that scaling $\hat{N}\propto c^{-4}$ as in (\ref{NRBS_Scaling2}) can only be regarded as a formal maneuver.  In other words, the strength of our extremal gravitational source can only be dialed in discrete steps, and certainly cannot reach below $\hat{N}= 1$. 

\section{Longitudinal RR Black Branes}
\label{longsec}

We now aim to repeat the analysis of the previous section for a different prototypical system: the longitudinal RR-charged black brane. The main steps and results will be the same as before, so we will be brief. 

More precisely, due to the requirement of positive fundamental string winding, we are interested in the extremal curved background sourced by a bound state of $\hat{K}$ D$p$-branes and $\hat{N}$ F1s, extended along the longitudinal spatial direction $\hat{x}$.  This is equivalent to $\hat{K}$ D$p$-branes that carry  $\hat{N}$ units of electric flux on their worldvolume. 
Within relativistic string theory, the resulting background was worked out long ago in \cite{Schwarz:1995dk,Russo:1996if,Costa:1996zd,Lu:1999uca}, and the string NR limit $c\to\infty$ with scaling (\ref{NRBS_Scaling}), holding $\hat{K}=K$ and $\hat{N}=N$ fixed, was taken in \cite{Gopakumar:2000na,Harmark:2000wv}. In the notation of \cite{Danielsson:2000mu,Guijosa:2023qym},\footnote{Save for the change of convention explained in footnote \ref{translationfoot}.} the result is the NR black $p$-brane background
\begin{subequations}
\begin{align}
ds^2 &= \mathsf{H}^{\frac{1}{2}}\frac{K^2}{g_s^2\nu^2}\frac{r^{7-p}}{L_{\mbox{\tiny long}}^{7-p}}\left( -dt^2+dx^2  \right) +\mathsf{H}^{-\frac{1}{2}}\left( dx_2^2+\dots +dx_p^2  \right) \nonumber \\ 
&   \qquad+\mathsf{H}^{\frac{1}{2}}\left( dx_{p+1}^2+\dots + dx_9^2  \right)~,\label{fondolong}\\
B&=\frac{1}{g_s^2 l_s^2}\frac{K^2}{\nu^2}\frac{r^{7-p}}{L_{\mbox{\tiny long}}^{7-p}}dt\wedge dx~, 
\label{fondoBlong}
\\
\exp(2\Phi)&=\frac{K^2}{\nu^2}\mathsf{H}^{\frac{5-p}{2}}
\frac{r^{7-p}}{L_{\mbox{\tiny long}}^{7-p}}~,
\label{fondophilong}
\end{align}
\label{longDp}
\end{subequations}
where
\begin{subequations}
\begin{align}
\mathsf{H}&=1+\frac{L_{\mbox{\tiny long}}^{7-p}}{r^{7-p}}~,
\\
r^2&\equiv x_{p+1}^2+\dots +x_{9}^2~,
\label{harmoniclong}\\
L_{\mbox{\tiny long}}^{7-p}&\equiv \frac{1}{(7-p)}\frac{K^2}{\nu}\frac{(2\pi)^{7-p}l_s^{\,7-p}}{\Omega_{8-p}}~,
\end{align}
\label{HRD}
\end{subequations}
with 
\begin{equation}
    \Omega_{8-p}=\frac{2\pi^{(9-p)/2}}{\Gamma\left(\frac{9-p}{2}\right)}
    \label{Omega}
\end{equation}
the area of the unit $(8-p)$-dimensional sphere, and 
\begin{equation}
    \nu\equiv \frac{N l_s^{\,p-1}}{ R_2\cdots R_p}
\end{equation}
the density of F1 charge bound the the $p$-brane. 
The appearance of this quantity in (\ref{longDp}) is associated with the fact that the effective tension in NR string theory for a longitudinal D$p$-brane carrying F1 density $\nu$ takes the form \cite{Danielsson:2000gi,Danielsson:2000mu}
\begin{equation}
T^{\,\mbox{\tiny long}}_{\mbox{\tiny D}p,\nu}=\frac{1}{2(2\pi)^p\nu g_s^2 l_s^{\,p+1}}~.
\end{equation}

Aside from the NS-NS fields (\ref{longDp}), the background includes nontrivial profiles for two RR fields: the usual ($p+2)$-form field strength $F_{(p+2)}$ sourced by the $K$ D$p$-branes \cite{Horowitz:1991cd,Polchinski:1995mt}, and the $p$-form field strength $F_{(p)}$ due to the well-known fact that the Chern-Simons couplings on the worldvolume of D$p$-branes with an electric field induce D$(p-2)$ charge \cite{Douglas:1995bn}.  

Just like in the black string example of the previous section, we see that the background (\ref{longDp}) does not have a proper SNC structure. Instead, it is a solution of the full relativistic Type IIA/IIB supergravity equations of motion, as laid out, e.g., in \cite{Polchinski:1998rr,Peet:2000hn,Itsios:2012dc}, and becomes nonrelativistic only asymptotically, as shown already in \cite{Danielsson:2000mu}. 

To confirm the proposal of \cite{Danielsson:2000mu} that (\ref{longDp}) is nothing more and nothing less than the curved background sourced in NR string theory by a stack of $K$ longitudinal D$p$-branes with $N$ units of F1 winding, an explicit calculation was carried out very recently in \cite{Guijosa:2023qym}, where the mere insertion of a boundary in the Gomis-Ooguri worldsheet action \cite{Gomis:2000bd}, with  boundary conditions appropriate for longitudinal D-branes, was found to give rise to expectation values for the massless closed string modes that correctly reproduce the $r\gg L_{\mbox{\tiny long}}$ behavior of (\ref{longDp}). The results show very clearly that longitudinal D-branes induce a $\lambda\bar\lambda$ term, which is consistent with the relativistic character of the background, and drives home the point that, from the perspective of the full NR string theory, it is unnatural to forbid this deformation.

Altogether then, we find again that post-limit the source is still strong enough to distort the background pronouncedly, generating a relativistic throat that connects to an asymptotically flat SNC region. As explained in \cite{Danielsson:2000mu,Guijosa:2023qym}, (i) the relation between the black $p$-brane (\ref{longDp}) and the stack of D$p$-branes embedded in flat SNC spacetime is entirely analogous to the standard open/closed string duality of relativistic string theory \cite{Polchinski:1995mt}; (ii) the property that only positively-wound strings can escape the throat corresponds to the fact that such strings are the only ones that can exist outside the D$p$-brane stack; and (iii) the physics of unwound closed strings in the throat of the black $p$-brane is dual to that of the (NCOS \cite{Seiberg:2000ms,Gopakumar:2000na}) open strings that describe excitations of the stack, with a translation between the two descriptions that is expected to be similar to the one discussed in \cite{Danielsson:2000ze,Amador:2002is,Amador:2003ju}. 


\section{Transverse RR Black Branes}
\label{transsec}
Another interesting example where we can explore the same questions is the extremal transverse RR-charged black $p$-brane. As explained in \cite{Danielsson:2000gi}, the microscopic source for it, a transverse D-brane, happens to have a tension that in the string NR limit is finite all by itself, and has the same form as in the parent theory,
\begin{equation}
T^{\,\mbox{\tiny trans}}_{\mbox{\tiny D}p}=\frac{1}{(2\pi)^p g_s l_s^{\,p+1}}~.
\end{equation}
It thus constitutes an exception to the generic requirement of positive fundamental string winding, similar to that of Newtonian gravitons, photons and D-brane collective coordinates \cite{Danielsson:2000mu}. In contrast with longitudinal D-branes, which carry string winding within themselves in the form of electric flux, transverse D-branes can only carry F1 winding through the open strings that describe their excitations, which indeed have a spectrum essentially identical to that of closed strings, implying in particular that on shell they must be positively wound  \cite{Danielsson:2000mu}.

We could thus choose to pile up, in the parent relativistic string theory, a macroscopic number of transverse D$p$-branes with  accompanying wound strings, which would generate a curved background of the D$p$-F1 junction type described in \cite{DHoker:2007zhm,DHoker:2007mci}. This would involve a nontrivial profile of the Kalb-Ramond field similar to the one encountered in the previous two sections, and we therefore expect it to lead again, in the NR limit, to a background that is globally relativistic but asymptotically SNC.

We will opt instead for examining a system of $K$ coincident transverse D-branes with no accompanying wound strings, or with at most a microscopic number of such strings. This is described gravitationally by the standard black $p$-brane solution \cite{Horowitz:1991cd}. This setup as well as its string NR limit were presented in \cite{Guijosa:2023qym}. For our purposes here, we will need to go back to the starting point, given by the background 
\begin{align}
    d\hat{s}^2&= \mathsf{H}^{-1/2}\left(-d\hat{t}^2+d\hat{x}_2^2+\ldots+d\hat{x}_{p+1}^2\right)
    +\mathsf{H}^{1/2}\left(d\hat{x}_1^2+d\hat{x}_{p+2}^2+\ldots+d\hat{x}_9^2\right)~,
\nonumber\\
\exp(2\hat{\Phi})&=\hat{g}_s^2\, \mathsf{H}^{\frac{3-p}{2}}~,
\label{blackpbrane}\\
\mathsf{H}&=1+ \sum_{m=-\infty}^{\infty}\frac{\hat{L}_{\mbox{\tiny trans}}^{\,7-p}}{\left[\left(\hat{x}_1-2\pi m \hat{R}\right)^2+\hat{r}_{\perp}^2\right]^{(7-p)/2}}~,
\nonumber\\
\hat{L}_{\mbox{\tiny trans}}^{\,7-p}&\equiv (4\pi)^{\frac{5-p}{2}}\Gamma\left(\frac{7-p}{2}\right)\hat{K} \hat{g}_s \hat{l}_s^{\,7-p}~,
\nonumber\\
\hat{r}_{\perp}^2&\equiv \hat{x}_{p+2}^2+\ldots+\hat{x}_9^2~,
\nonumber
\end{align}
supplemented with the usual RR $(p+2)$-form field strength $F_{(p+2)}$. 
To reflect the fact that the solution is localized at the origin of the compact longitudinal spatial direction $\hat{x}_1\simeq \hat{x}_1 + 2\pi \hat{R}$, we have taken the harmonic function $\mathsf{H}$ to be a sum of terms that incorporate mirror images at $\hat{x}^1=2\pi m \hat{R}$ for all integer $m$.

Additionally, in anticipation of the limit that will take us to NR string theory, without affecting (\ref{blackpbrane}) we turn on a constant critical Kalb-Ramond field
\begin{equation}
    \hat{B}=\left(1-\frac{\mu}{c^2}\right)d\hat{t}\wedge d\hat{x}~, 
\label{criticalB}
\end{equation}
with $\mu$ and arbitrary constant, that can be set to zero
\cite{Danielsson:2000mu,Guijosa:2023qym}.
We wish to subject (\ref{blackpbrane}) and (\ref{criticalB}) to the string NR limit $c\to\infty$, with
\begin{equation}
\hat{l}_s=l_s~,\: 
\hat{g}_{s}=c\,g_{s}~, \:
\hat{K}=K~, \:
\hat{R}=cR~,\,
\hat{t}=c\,t~,\, 
\hat{x}=c\,x~, \quad 
\hat{x}_2=x_2~, \ldots,\,
\hat{x}_{9}=x_9~.
\label{NRBS_Scaling3}
\end{equation}
Up to corrections of $\cO(c^{-(6-p)})$, this has the effect \cite{Guijosa:2023qym} of collapsing the harmonic function $\mathsf{H}$ to be nontrivial only on the 8-dimensional plane $x=2\pi m R\,$:
\begin{align}
\mathsf{H} &= 1+\frac{L_{\mbox{\tiny trans}}^{7-p}}{r_\perp^{6-p}}\sum_{m=-\infty}^{\infty}
\delta\!\left(x^1-2\pi mR\right)~,
\label{deltafcn}\\
L_{\mbox{\tiny trans}}^{7-p}&\equiv 2^{5-p}\pi^{\frac{6-p}{2}}\,\Gamma\!\left(\frac{6-p}{2}\right)K g_s l_s^{\,7-p}~.
\nonumber
\end{align}
Because of this localization, we can revert to considering $x$ only within the compact range $[0,2\pi)$, and therefore write the delta function simply as 
$\delta(x^1)$.
Using (\ref{deltafcn}), we see that (\ref{blackpbrane}) and (\ref{criticalB}) read
\begin{align}
    d\hat{s}^2&= c^2 \left(-\mathsf{H}^{-1/2}dt^2+\mathsf{H}^{1/2}dx^2\right)
\nonumber\\
  &\quad +\mathsf{H}^{-1/2}\left(dx_2^2+\ldots+dx_{p+1}^2\right)
    +\mathsf{H}^{1/2}\left(dx_{p+2}^2+\ldots+dx_9^2\right)~,
\nonumber\\
\hat{\Phi}&=\ln c +\ln\left(g_s\, \mathsf{H}^{\frac{3-p}{4}}\right)~,
\label{transverseDp}\\
\hat{B}&=\left(c^2-\mu\right)dt\wedge d x~. 
\nonumber
\end{align}
Comparing this with (\ref{NRExpansion_GBPHi}), we can successfully read off
\begin{equation}
\begin{aligned}
& \tau_{\mu}^{0}=\mathsf{H}^{-1/4}\delta^0_{\mu}~, \quad
\tau_{\mu}^{1}=\mathsf{H}^{1/4}\delta^1_{\mu}~, 
\\
E_{\mu\scriptstyle ||}^{A'}&=\mathsf{H}^{-1/4}\delta_{\mu\scriptstyle ||}^{A'}~, \quad 
E_{\mu\scriptstyle \perp}^{A'}=\mathsf{H}^{1/4}\delta_{\mu\scriptstyle \perp}^{A'}~,
\\
m_{\mu}^{A}&=C_{\mu}^{A}=0~,
\\
\Phi&=\ln\left(g_{s}\,\mathsf{H}^{\frac{3-p}{4}}\right)~,
\\
B&=-\mu\, dt\wedge dx~,
\end{aligned}
\label{yessnc}
\end{equation}
where the subindices $\mu\scriptstyle ||$ and
$\mu\scriptstyle\perp$
denote the transverse directions that are respectively parallel ($2,\ldots,p+1$) and perpendicular ($p+2,\ldots,9$) to the brane. 

We conclude then that, unlike the black string and the longitudinal black brane, the transverse black brane \emph{does have a proper SNC structure}, which is flat almost everywhere, with the exception of the 8-dimensional plane $x=0$. Within that plane, it becomes flat as $r_{\scriptstyle\perp}\to\infty$.

\section*{Acknowledgements}
We thank J.~Antonio Garc\'{\i}a for useful conversations. Our work was partially supported by Mexico's National Council of Science and Technology (CONACyT) grant A1-S-22886 and DGAPA-UNAM grant IN116823. DA was additionally supported by a DGAPA-UNAM postdoctoral fellowship. 

\bibliography{asymptoticallynr}

\bibliographystyle{./utphys}

\end{document}